\documentclass[aps,twocolumn,reprint,superscriptaddress]{revtex4-1}%
\usepackage{float}
\usepackage{graphicx}
\usepackage{amsmath,amssymb,bbm,mathrsfs,bm,braket,color,graphicx,comment,amsfonts,dsfont}
\usepackage[colorlinks,citecolor=blue,urlcolor=blue]{hyperref}
\usepackage[mathscr]{euscript}
\usepackage[normalem]{ulem}
\usepackage{comment}

\DeclareMathAlphabet\mathbfcal{OMS}{cmsy}{b}{n}

\usepackage{extarrows} 
\usepackage{mathrsfs}

\begin{document}

\title{Bending nanoribbon to induce large anisotropic magnetoconductance
} 

\author{Ponder Liu}
\affiliation{Department of Physics, National Cheng Kung University, Taiwan}
\affiliation{Center for Quantum Frontiers of Research and Technology (QFort), National Cheng Kung University, Tainan 70101, Taiwan}

\author{Hao-Cheng Hung}
\affiliation{Department of Physics, National Cheng Kung University, Taiwan}
\affiliation{Center for Quantum Frontiers of Research and Technology (QFort), National Cheng Kung University, Tainan 70101, Taiwan}

\author{You-Ting Huang}
\affiliation{Department of Physics, National Cheng Kung University, Taiwan}
\affiliation{Center for Quantum Frontiers of Research and Technology (QFort), National Cheng Kung University, Tainan 70101, Taiwan}
\affiliation{Research Center for Applied Sciences, Academia Sinica, Taipei 11529, Taiwan}

\author{ Jia-Cheng Li }
\affiliation{Center for Quantum Frontiers of Research and Technology (QFort), National Cheng Kung University, Tainan 70101, Taiwan}
\affiliation{Program on Key Materials, Academy of Innovative Semiconductor and Sustainable Manufacturing, National Cheng Kung University, Tainan 70101, Taiwan}

\author{Carmine Ortix}
\affiliation{Dipartimento di Fisica ``E. R. Caianiello'', Universit\`a di Salerno I-84084 Fisciano (Salerno), Italy}

\author{Ching-Hao Chang}
\affiliation{Department of Physics, National Cheng Kung University, Taiwan}
\affiliation{Center for Quantum Frontiers of Research and Technology (QFort), National Cheng Kung University, Tainan 70101, Taiwan}
\affiliation{Program on Key Materials, Academy of Innovative Semiconductor and Sustainable Manufacturing, National Cheng Kung University, Tainan 70101, Taiwan}

\begin{abstract}

When a nanoribbon is bent under a homogeneous external magnetic field, the effective magnetic field inside becomes either homogeneous or inhomogeneous, depending on the direction of the field. This enables the selective creation of bulk, interface, and edge magnetic states in the bent structure, for a magnetic field with a strength.  We establish theoretically that these tuneable states lead to a strong geometry-induced anisotropic magnetoconductance (GAMC) in perpendicularly bent nanoribbon, which can reach up to 100\%. Moreover, the GAMC can be further enhanced to 200\%, 300\%, or even higher by either further bending or tuning the bending angle. The potential of this phenomenon for practical applications is demonstrated by its stable anisotropy, which remains consistent across a wide range of Fermi energies, can be observed even at weak magnetic fields and room temperature, and occurs in various systems such as two-dimensional electron gas (2DEG) and graphene.
\end{abstract}

\maketitle
\textcolor{blue}{\textit{Introduction}} --- 
Since the successful isolation of graphene in 2004 \cite{graphene2004}, research into two-dimensional (2D) materials has gained significant momentum. Unlike conventional bulk materials, where properties are dominated by bulk states, 2D materials are primarily governed by surface or interface states that are artificially controllable. This distinction leads to unique electrical, optical, mechanical, and thermodynamic properties, making them a rich field of study.  Additionally, the emergence of special geometric structures, such as curves, helices, and spirals, in 2D materials opens up new possibilities for tailoring their properties \cite{xie2009,ma2020,wang2024,shen2019}. For instance, the unconventional superconductivity can be created by twisting bilayer graphene \cite{Nature_SC}. Magneto-transport-like phenomena, such as nonlinear Hall effects, can exhibit in a corrugated graphene system with time reversal symmetry 
\cite{PhysRevLett.115.216806, Nature.volume.565, Nature.Reviews.3, Nature.Electronics.2021.4}.

Building on advances in geometry-induced phenomena, the anisotropic magnetoconductance (AMC) effect presents an intriguing avenue for exploration in 2D materials. 
The variation of electric current with the magnetization direction is typically linked to intrinsic properties such as spin-orbit coupling and the scattering behavior of charge carriers \cite{SMIT1951612, PhysRevLett.94.127203}. This phenomenon has been experimentally observed in materials like Fe, Co, Ni, and their alloys, with practical applications in devices such as magnetic field sensors \cite{mcg1975, amr-book}. 
This raises the question of whether the AMC effect, traditionally dependent on intrinsic material properties, could be tuned through geometric manipulation in 2D materials \cite{PhysRevLett.113.227205,Nano.Letters.2017.5,Nature.Electronics.2022.5}.

\begin{figure}[h!]
    \centering
    \includegraphics[width=.95\linewidth]{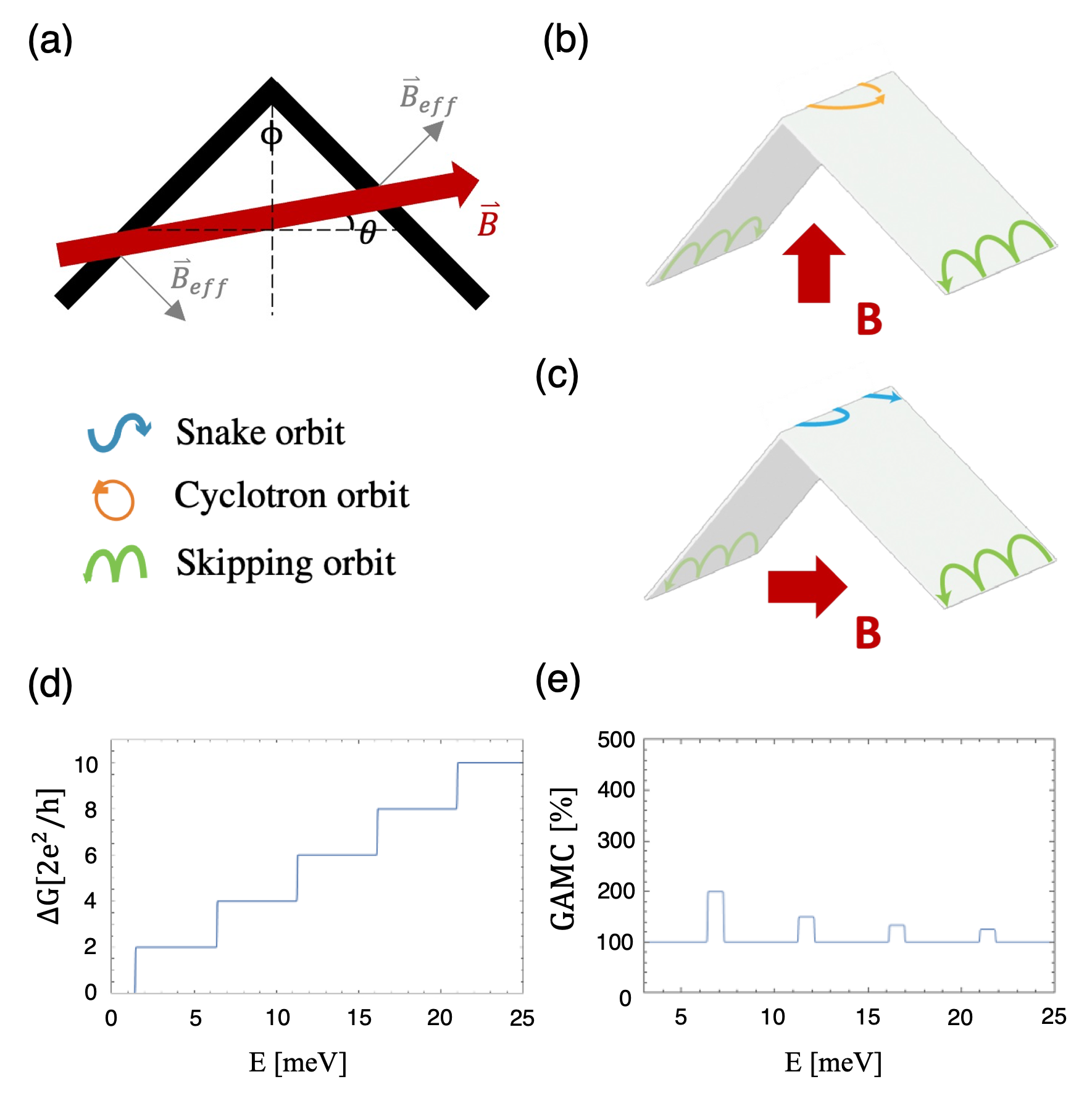}
    \caption{(color online). (a), 
    Front view of the bent 2DEG nanoribbon with bending angle $\phi=\pi/2$ subjected to a 4 T transverse magnetic field. 
    The semi-classical motion of charge carriers in the applied magnetic field with angle $\theta=\pi/2$ and $\theta=0$ are shown in panel (b) and (c), respectively. (d), The conductance difference between two systems increases as a function of Fermi energy. (e), The ballistic anisotropic magnetoconductance (GAMC) ratio as a function of Fermi energy.  Panels (d) and (e) are estimated at temperature T = 0.1 K.}
    \label{fig1}
\end{figure}

In this study, we theoretically demonstrate that geometry-induced anisotropic magnetoconductance (GAMC) can reach 100\% or even 300\% in the electronic transport of well-developed bent nanomembranes \cite{mcg1975, amr-book}, as is shown in Fig.1. Unlike the AMC observed in specific traditional magnetic materials, the GAMC we propose is entirely driven by the bent geometry, which induces a transition between magnetic states when the direction of the magnetic field is switched (see Figs.1(b) and (c)). Therefore, this effect is widely applicable to conducting 2D nanomembranes, such as 2DEG and graphene, which host different types of charge carriers. Moreover, the GAMC value in bent graphene remains fixed over a wide range of Fermi energies, persists at room temperature, and can be realized in weak magnetic fields ($<$1T), offering potential for expanding AMC applications into novel sensing or control functionalities \cite{AMC_application}. 

\textcolor{blue}{\textit{Two-dimensional electron gas}} --- 
We begin by considering a two-dimensional electron gas (2DEG) system \cite{2DEG1,2DEG2} formed in a GaAs/AlGaAs heterostructure. The carriers therein have effective mass $m^\star=0.067\,m_e$ \cite{BEENAKKER19911}, and exhibit mean free path up to $10\mu m$ at low temperature \cite{D.Weiss_1989}.
The system can be fabricated into bent nanostructures through selective underetching of pseudomorphically strained bilayers. 
In our model, the bending corners are modeled by quarter-circles occupying $1\%$ of the total length $L$.

For the 2DEG nanoribbon with bending angle $\phi$ and is subjected to an homogeneous magnetic field with strength $B_0$ (see  Fig.1 (a)), the effective magnetic field on the nanoribbon is 
\begin{align}
\notag & B_{\rm eff,L}=-B_0\cos(\theta+\phi),\\
 & B_{\rm eff,R}=B_0\cos(\theta-\phi)
\end{align}
where L and R indicates the left and right halves of the nanoribbon, respectively, $\theta$ is the angle of applied magnetic field, and $\phi$ is the bending angle of nanoribbon.

For $\theta=\pi/2$ and $\phi=\pi/2$ in Eq.(1), the effective magnetic field throughout the ribbon becomes uniform with a magnitude of $B_0/\sqrt{2}$. The carriers form localized cyclotron orbits in the central region of the ribbon and skipping orbits at two edges when the field is strong enough to reduce the diameter of cyclotron orbit shorter than the ribbon width (see Fig.1 (b)). However, when $\theta$ is changed to zero, the effective magnetic field becomes a magnetic-field dipole. This leads to carriers in the left half experience an effective field of $-B_0/\sqrt{2}$, while in the right half experience an effective field of $B_0/\sqrt{2}$. As a result, the formation of magnetic-field dipole results in two kinds of distinct magnetic states coexisting -- carriers in the central region form snake states, which transport along the longitudinal direction, and form skipping states at edges that propagate in a direction opposite to the snake states (see Fig.1 (c)).

It can be expected that the variation in species of magnetic states (see Fig.1 (b) and Fig.1 (c)) may drive the anisotropic magnetotransport effect. To investigate this, we model a nanoribbon system with a
width $314 \text{nm}$ and the magnetic field magnitude is fixed at $B_0 = 4 \text{T}$. This setup ensures that the magnetic length $l_B = \sqrt{\hbar/eB}$ is much smaller than the system width. 
Under these conditions, the Zeeman splitting and other spin-related effects are negligible, allowing us to focus solely on orbital contributions (see Supplemental Material).
To quantify this magnetotransport effect, we define the GAMC  magnitude as \cite{PhysRevLett.94.127203}
\begin{align}
{\rm GAMC }(\theta) = \frac{G(\theta)-G(\pi/2)}{G(\pi/2)}
\end{align}
The variations of distinct magnetic states in different B-field direction, which, in turn, lead to corresponding difference in the longitudinal conductance of the system and further result in significant GAMC magnitude up to 100\%, as shown in Figs.1(d) and (e).

\begin{figure}[h!]
    \centering
    \includegraphics[width=.95\linewidth]{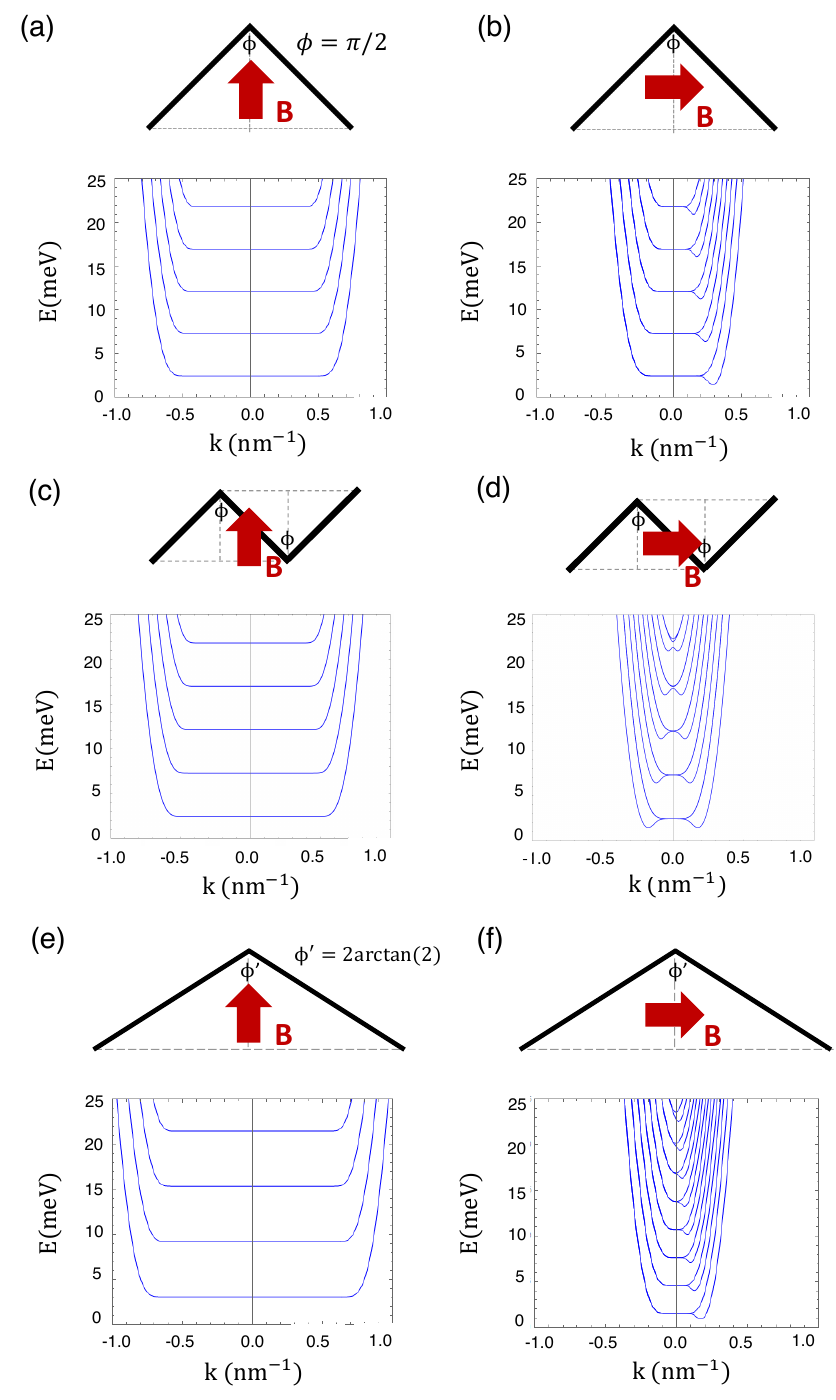}
    \caption{(color online). Magnetic spectrum of the nanoribbons with width $L=32\pi^2nm$ with different bending shape subjected to a 4 T magnetic field with relative angles $\theta=\pi/2$ and $0$. (a) and (b) are with bending angle $\phi=\pi/2$. (c) and (d) are bent twice with $\phi=\pi/2$. (e) and (f) are bent once with $\phi'=2\arctan(2)$.}
    \label{fig2}
\end{figure}

The GAMC shown in Fig.1 (e) is estimated from the magnetic band structures that computed using the diagonalization approach (see Supplemental Material).
In this calculation, the effects of Zeeman splitting and the effective potential induced by the bent geometry \cite{costa1981} are neglected,
as their influence on the band structure is perturbative and negligible (detailed discussion is presented in Supplemental Material).
To understand the correspondence between the species of magnetic states and the conductance and further manipulate the GAMC, we compute the magnetic band structures of the system in Fig.1 and other two 2DEG nanomembrane with different bent geometry structure. Figures 2 (a) and (b) illustrate the system of Figs.1, specifically, a nanoribbon bent once with a bending angle of $\phi=\pi/2$. In contrast, Figs.2 (c) and (d) depict a system bent twice at the same bending angle, whereas Figs.2 (e) and (f) correspond to nanoribbons bent once with a bending angle of $\phi=2\arctan2$ instead of $\pi/2$. 

For Figs.2 (a), (c), and (e), the ribbons experience uniform effective magnetic fields. Consequently, their band structures comprise Landau levels, along with sloped bands arising from finite boundary effects at large momentum. On the other hand, in Figs.2 (b), (d), and (f), the ribbons experience magnetic-fields dipole. In these cases, electrons form snake states around the interfaces where the effective magnetic field directions change, while cyclotron states with smaller cyclotron radii are observed in small regions near the centers of each halves, where the effective magnetic fields are uniform. The small cyclotron states correspond to the double degenerate Landau level bands near $k=0$ (Fig.2 (d) is triple degenerate because there are three cyclotron states in the system). The edge and snake states manifest as degenerate and non-degenerate sloped bands at large momentum respectively.

Notably, according to Eq.(1), the magnitude of the effective magnetic field
$\left| B_{\rm eff}\right|$ is $B_0/\sqrt{2}$ in Figs.2(a)–(d), while in Figs.2(e) and (f), it is $2B_0/\sqrt{5}$ and $B_0/\sqrt{5}$, respectively. This difference arises from the choice of bending angle $\phi = 2\arctan{2}$ for Figs.2(e) and (f) instead of $\pi/2$ used in Figs.2(a)–(d). Given that the energy gap of Landau level is determined by $\Delta E = \hbar q\left|B_{\rm eff}\right|/m^\star$,
and that $\left| B_{\rm eff} \right|$ in Fig.2 (e) is twice that in Fig.2 (f), the corresponding energy gap in Fig.2 (e) is consequently twice as large as in Fig.2 (f).

Using the band structures obtained in Fig.2, the conductance of the bent system can be readily calculated \cite{PhysRevLett.113.227205}, allowing for the determination of the GAMC in Eq.(2).
The longitudinal conductance is given by the Landauer formula in ballistic transport $G = N(2e^2/h)$, where the factor of 2 accounts for the electron spin degeneracy, and $N$ represents the number of open conducting channels, i.e., the number of intersections between the Fermi level and bands \cite{10.1063/1.531590}.
To incorporate the effects of temperature, we introduce the Fermi-Dirac distribution:
\[f(E,T)=\frac{1}{e^{(E-E_f)/k_BT}+1}\]
where $E$ is the energy, $E_f$ is the Fermi energy, $k_B$ is the Boltzmann constant, and $T$ is the temperature.

Figure 3 illustrates the GAMC of three systems at various temperatures. For the first system ($\phi = \pi/2$), the GAMC  is approximately 100\%, and for the second system (bent twice), the GAMC increases to about 200\%. In the third system ($\phi = 2\arctan(2)$), the GAMC rises from 100\% to roughly 300\% compared to the first system. This enhancement arises because the energy gap in Fig. 2(e) is twice as large as that in Fig. 2(f).
As the temperature rises, the GAMC decrease slightly but remain significantly higher than in the conventional AMC effect. At room temperature, the GAMC for the three systems are reduced to 88\%, 131\%, and 190\%, respectively, demonstrating the robustness of the GAMC effect even under thermal fluctuations (see Supplemental Material). 

\begin{figure}[h!]
    \centering
    \includegraphics[width=0.8\linewidth]{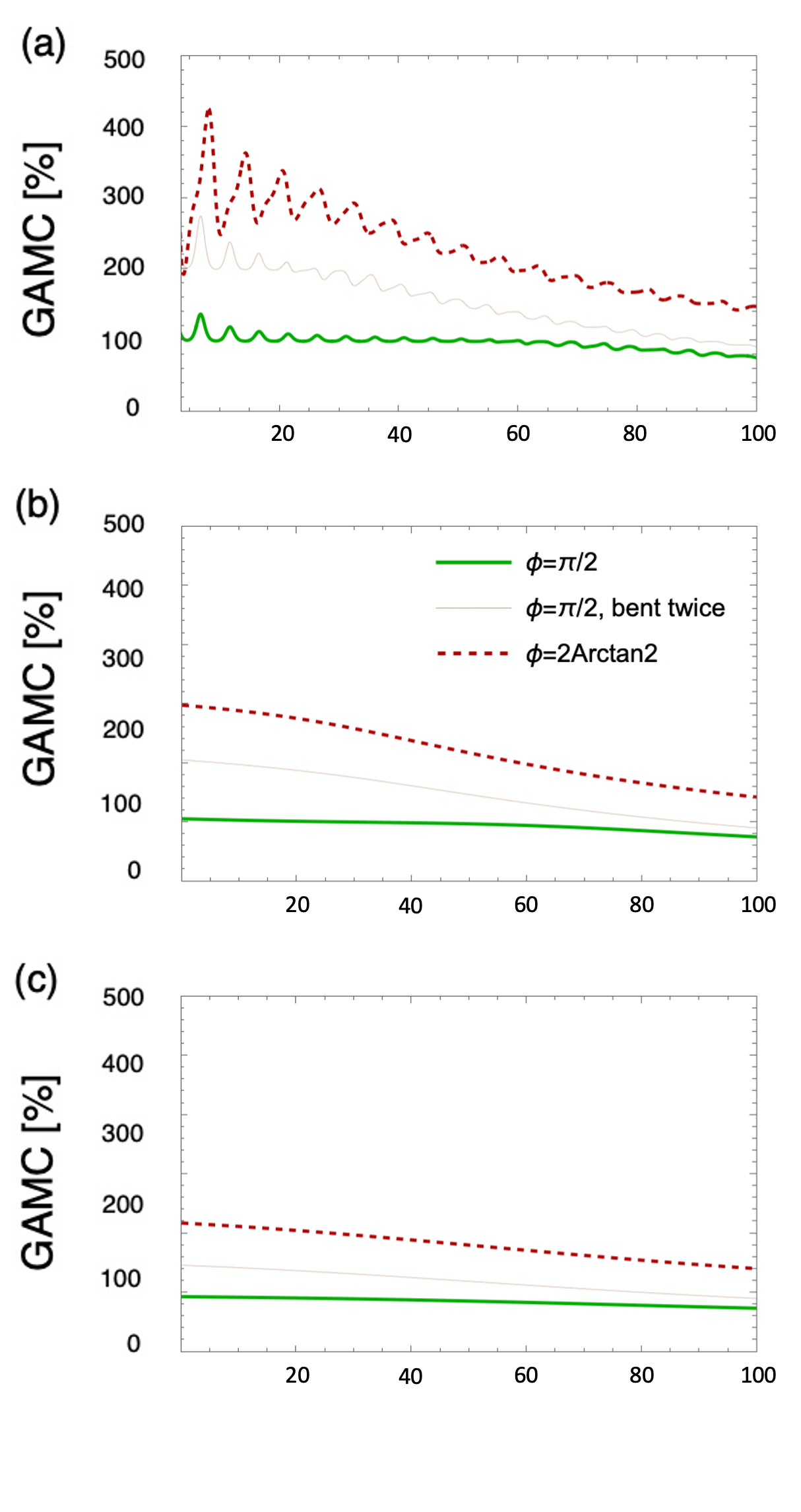}
    \caption{(color online). (a) GAMC  ratio in T=5K as a function of Fermi energy. (b) Same for T=100K. (c) Same for T=300K.}
    \label{fig3}
\end{figure}

\textcolor{blue}{\textit{Declining of GAMC at high Fermi energy}}---
The GAMC begins to drop significantly when the Fermi energy reaches a critical value at which the diameter of electron cyclotron orbits becomes comparable to the characteristic geometric width of the system, namely the width of nanoribbon in our proposed system. At this point, the distinction between magnetic states such as skipping and snake orbits becomes indistinguishable. This critical value is called the cut-off energy $E_c$.
To determine $E_c$, we analyze the interplay between two length scales: the geometry length $L'$, which is the width of each halves, and the diameter of electron cyclotron orbits $2R_c=2\frac{\hbar k_c}{qB}$, where the wavevector $k_c$ is related to $E_c$ by $k_c=\sqrt{2 m^\star E_c}/\hbar$. The value of $E_c$ is derived as:

\begin{align}
E_c = \frac{1}{4} q v_F B_{\rm eff}~L',
\label{eq_Ec}
\end{align}

where the Fermi velocity $v_F=qB_{\rm eff}L'/(2m^\star)$
is respected to the magnetic-orbit diameter $2R_c$ being the same as the geometry length $L'$ (see Supplemental Material).
For $E_F>E_c$ (so that $2R_c>L'$), the GAMC effect contributed by skipping and snake orbits reduces dramatically because these two magnetic states are eliminated.
For the three systems considered in Figs. 2 and 3, the estimated $E_c$ values are 64.6 meV, 28.7 meV, and 25.8 meV, respectively. 
As shown in Fig.3, the GAMC decrease once the Fermi energy $E_F$ exceeds the corresponding $E_c$ for each system, indicating the diminishing impact of magnetic state distinctions under these conditions.

\textcolor{blue}{\textit{Graphene}} --- 
graphene, in this regards, offers a significant advantage in increasing the critical energy $E_c$, as the Fermi velocity in graphene ($v_F \approx 10^6 m/s$ \cite{RevModPhys.81.109}) is an order of magnitude greater than in 2DEG ($v_F \approx 2.7 \times 10^5 m/s$ at $E = 14$ meV \cite{BEENAKKER19911}), and the mean free path can extend to $28\mu m$ at low temperature and remains above $1\mu m$ at 200K \cite{doi:10.1021/acs.nanolett.5b04840}. The $E_c$ in graphene is thus expected to exceed that in 2DEG systems by an order of magnitude.
As shown in Fig.4 (a), the 100\% GAMC persists over much broader ranges of Fermi energies for both armchair and zigzag graphene nanoribbons compared to 2DEG systems(Our numerical simulations of graphene nanoribbons are based on the kwant package \cite{Groth2014Jun}. Detailed calculation methods are provided in Supplemental Material). This extended GAMC plateau highlights the superior capability of graphene in sustaining a robust ballistic GAMC effect over a wider operational energy range.

\begin{figure}
\centering
\includegraphics[width=.9\linewidth]{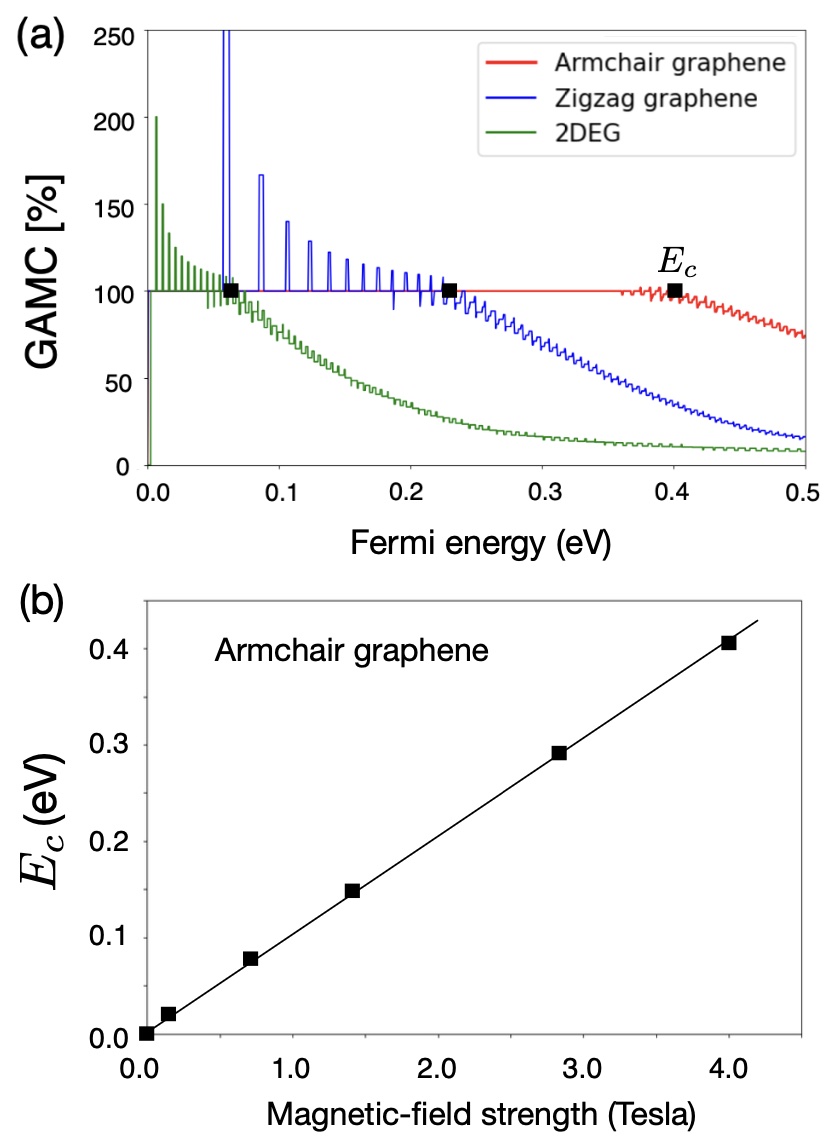}
    \caption{(color line). (a) The GAMC  ratio of armchair, zigzag graphene, and 2DEG
nanoribbon with length $L=32\pi^2nm$, subjected to magnetic field $B=4T$ as a function of Fermi energy. (b)The energy cut-off $E_c$ of the GAMC effect varys with the magnetic field strength in the graphene nanoribbon. The value of $E_c$ is obtained numerically (see Supplemental Material).}
    \label{fig4}
\end{figure}

From another perspective in Eq.(3), the GAMC effect in a graphene system can be achieved within the same energy range as in 2DEG systems by reducing the magnetic field strength by an order of magnitude. As shown in Fig.4 (b), for instance, applying a weak magnetic field (0.5 Tesla) to a graphene system generates a GAMC effect comparable in energy range (0.1 eV) to that produced in a 2DEG system under higher field strength (4 Tesla). The results in Fig.4 indicate that the GAMC effect can be extended to other conductive materials, and its cut-off energy $E_c$, as determined in Eq.(3), can be tuned artificially either by choosing the nanoribbon width $L$ or by selecting different 2D material species with different Fermi velocity $v_F$, enabling flexible design and optimization of GAMC-based systems.

\textcolor{blue}{\textit{Summary}} --- 
We have predicted the existence of a tunable geometry-induced anisotropic magnetoconductance— a phenomenon where the ballistic conductance changes by several hundred percent depending on the direction of an applied magnetic field. This effect occurs in bending nanoribbon systems, including both conventional non-magnetic semiconducting materials and graphene. The GAMC effect originates purely from the bent geometry of the nanostructures, allowing its value to be tuned by adjusting the bending angle or increasing the number of bends in the conducting nanomembrane (analysis of multiple-bent nanomembranes is provided in Supplemental Material).

It is worth noting that the experimental realization of the predicted GAMC depends on the relative magnitudes of three energy scales: thermal energy \(k_B T\), Fermi energy \(E_F\), and the cut-off energy \(E_c\), which depends on the material. Since \(E_c\) is approximately 64.6\,meV for the 2DEG system (Fig.3) and 200\,meV for the graphene system (Fig.4), we expect the GAMC effect to be observable at room temperature (\(k_B T \approx 25\,\text{meV}\)) in the ballistic conducting nanoribbons with Fermi energies smaller than \(E_c\).

\section*{Author contribution} P. L., H.-C.H. and Y.-T.H. contributed equally to this work.
Y.-T.H. (l26104115@gs.ncku.edu.tw) and C.-H.C. (cutygo@phys.ncku.edu.tw) are corresponding authors.

\section*{Acknowledgements} We acknowledge the financial support by the National Science and Technology Council (Grant numbers 112-2112-M-006-026-, and 113-2112-M-006-08 ) and National Center for High-performance Computing for providing computational and storage resources. C.O. ac- knowledges support from the MAECI project “ULTRAQMAT”. This work was supported in part by the Higher Education Sprout Project, Ministry of Education to the Headquarters of University Advancement at the National Cheng Kung University (NCKU).

\bibliography{refs.bib}

\end{document}